\documentclass{article}

\usepackage{comment}
\usepackage{authblk}
\usepackage{graphicx}
\usepackage{amsmath}

\usepackage{xcolor}
\newcommand{\red}[1]{{#1}}

\usepackage{lineno}

\begin{document}

\title{Nanosecond gating of superconducting nanowire single-photon detectors using cryogenic bias circuitry}

\author[1,$\dagger$]{Thomas Hummel} 
\author[1]{Alex Widhalm} 
\author[1]{Jan Philipp Höpker} 
\author[1]{Klaus D. Jöns}
\author[2]{Jin Chang}
\author[3]{Andreas Fognini}
\author[4]{Stephan Steinhauer}
\author[4]{Val Zwiller} 
\author[1]{Artur Zrenner}  
\author[1,*]{Tim J. Bartley}

\affil[1]{Department of Physics, Paderborn University, Warburger Str. 100, 33098 Paderborn, Germany}
\affil[2]{Department of Quantum Nanoscience, Delft University of Technology, 2628 CJ, Delft, the Netherlands}
\affil[3]{Single Quantum B.V., 2629 HH, Delft, the Netherlands}
\affil[4]{Department of Applied Physics, Royal Institute of Technology (KTH), SE-106 91 Stockholm, Sweden}

\affil[$\dagger$]{thomas.hummel@upb.de}
\affil[*]{tim.bartley@upb.de}
\maketitle{}

\begin{abstract}
Superconducting nanowire single-photon detectors (SNSPDs) show near unity efficiency, low dark count rate, and short recovery time. 
Combining these characteristics with temporal control of SNSPDs broadens their applications as in active de-latching for higher dynamic range counting or temporal filtering for pump-probe spectroscopy or LiDAR. 
To that end, we demonstrate active gating of an SNSPD with a minimum off-to-on rise time of $2.4\,\mathrm{ns}$ and a total gate length of $5.0\,\mathrm{ns}$.
We show how the rise time depends on the inductance of the detector in combination with the control electronics.
The gate window is demonstrated to be fully and freely, electrically tunable up to $500\,\mathrm{ns}$ at a repetition rate of $1.0\,\mathrm{MHz}$, as well as ungated, free-running operation.
Control electronics to generate the gating are mounted on the $2.3\,\mathrm{K}$ stage of a closed-cycle sorption cryostat, while the detector is operated on the cold stage at $0.8\,\mathrm{K}$.
\red{We show that the efficiency and timing jitter of the detector is not altered during the on-time of the gating window.}
We exploit gated operation to demonstrate a method to increase in the photon counting dynamic range by a factor $11.2$, as well as temporal filtering of a strong pump in an emulated pump-probe experiment.
\end{abstract}

\section{Introduction}
The ability to actively switch an optical detector on and off enables noise reduction and temporal filtering. 
This is particularly crucial at the single-photon level for many applications such as quantum dot resonance fluorescence spectroscopy~\cite{Wu1975, Astafiev2010, Ding2016, Kirsanske2017, Scholl2019} and LiDAR measurements~\cite{Kim2012, Jantzi2021, Li2021} to temporally filter pump noise, or for mid-IR detectors~\cite{Chang2022} to reduce the dark count rate.
In this single-photon regime, the invention of the superconducting nanowire single-photon detectors (SNSPDs)~\cite{Goltsman2001} revolutionised single-photon detection due to their low noise, fast rise time response, low timing jitter, and high detection efficiency~\cite{Hadfield2009, Natarajan2012, Marsili2013, Shibata2015, Redaelli2016, Esmaeil2017, Nicolich2019, korzh2020, Reddy2020, ChangJ2021, EsmaeilZadeh2021}.
Despite this, active gating of SNSPDs has \red{up to now only} been \red{demonstrated down to }
hundreds of nanoseconds~\cite{Schuck2013,Liu2021}. 
Nevertheless, for other single-photon detectors, current benchmarks for gating are significantly shorter. 
For example, gated avalanche photo detectors (APDs) achieve rise times of $200\,\mathrm{ps}$ with timing jitters of $30\,\mathrm{ps}$~\cite{Boso2013}.
Thus the question arises how fast an SNSPD can be \red{deterministically gated and which applications this opens up}.

A particular challenge of working with SNSPDs is their cryogenic operating conditions. 
Maintaining Radio Frequency (RF) signal integrity at the SNSPD is a challenge to overcome since the control electronics and SNSPD are either connected through a coaxial cable from ambient to cryogenic temperature, or the control electronics operate under the same cryogenic conditions as the SNSPD.
A conventional method of overcoming this connection is by RF-biasing the SNSPD~\cite{Akhlaghi2012, Doerner2019, Knehr2020}, where the current is modulated at a fixed RF-frequency.
Independent control over the gate window requires deterministic gating~\cite{Schuck2013, Zhang2017, Liu2021} where a non sinusoidal current is fed through the SNSPD, however these methods are limited in their timing resolution.
Another approach in modifying the timing characteristics of an SNSPD comes from active quenching~\cite{Ravindran2020} of an SNSPD, where the bias current is actively switched off after a detection event.
Alternatively, temporal control can be achieved by modulating the optical signal with an Acousto-Optic Modulator (AOM), which can achieve GHz bandwidth\cite{Rashed2014,Zejie2021} but is limited by the optical losses~\cite{Savage2010,RamirezMelendez2017}

A key aspect in the manipulation of the temporal behaviour of the SNSPD lies in how fast the bias current through the SNSPD can be switched.
In this paper we investigate the effect of the basic electronic design on the on-and-off speed of an SNSPD in terms of its electrical properties and provide a simple model to predict the characteristic switching times. 
Further, we demonstrate deterministic gating of an SNSPD with a rise time of $2.4\,\mathrm{ns}$ and a smallest achievable gating window of $5\,\mathrm{ns}$, and compare the detector figures of merit to the free-running regime.
Finally we show proof of principle applications of a gated SNSPD, including temporal filtering, noise reduction, and exploiting photon statistics for high dynamic range photon counting.

\section{Electronic gating circuitry} \label{GatingCircuit}
Superconducting nanowire single-photon detectors are based on a narrow strip of superconducting material, cooled below its critical temperature.
The detector is biased close to its critical current, such that when a photon is absorbed the superconductivity breaks down~\cite{Goltsman2001, Natarajan2012}.
This induces a resistive area which causes the current to divert through a shunt resistor creating a voltage pulse which can be detected.
The probability that an absorbed photon causes the localised breakdown of superconductivity depends on the energy of the photon and the bias current through the SNSPD \cite{Marsili2012, Wolff2021}.
Modulating this bias current over time therefore allows modulation of the detection efficiency.
Realising pulsed modulation between no current and the current which maximises detection efficiency therefore allows a full on-off switching of an SNSPD, where deterministic gating can be achieved with control over the start time and pulse duration of the bias current.

\begin{figure}[ht!]
\centering\includegraphics{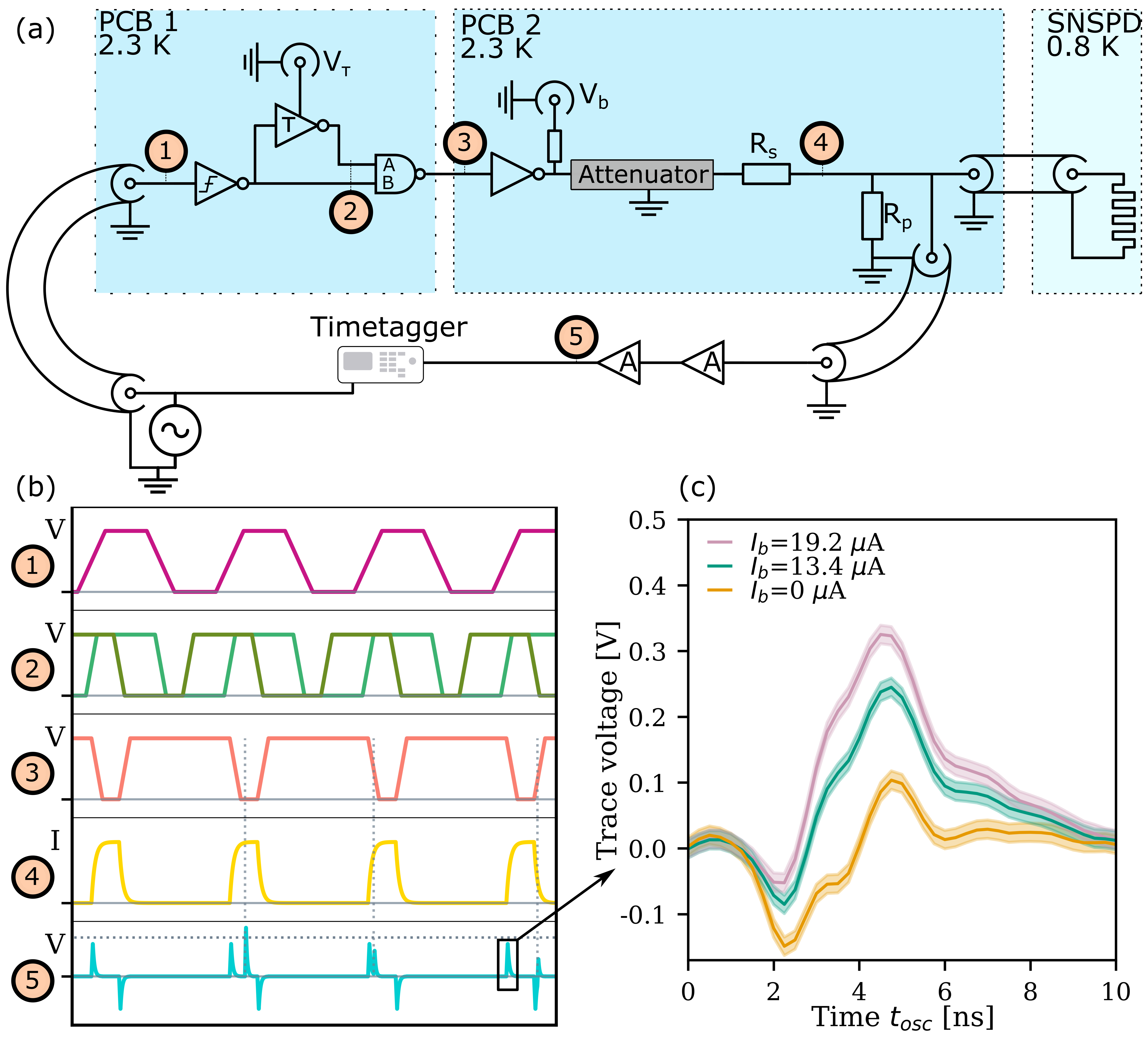}
\caption{Operating principle of the electrical circuit generating the gating signal. 
(a) Simplified model of the electrical circuit with relevant signals marked.
(b) Stylistic representation of the steps used to generate a gating signal where markers $1$ to $5$ refer to the electrical signal at points $1$ to $5$ in (a). 
An input pulse is inserted in the circuit (b$1$), which is buffered, and split in two paths.
The upper path is inverted and delayed with a tuneable delay while the lower path is unchanged (b$2$).
A NAND gate recombines the two paths (b$3$) and generates an inverted narrow pulse corresponding to the width of the induced time delay in the top path.
The narrow pulse is inverted again and sent through an attenuator circuit to transform it into a bias current (b$4$) through the SNSPD.
Dotted vertical lines indicate impinging photons showing respectively four different cases: no photon, photon impinging at the set current, photon impinging while the current is loading, and photon impinging while the current is unloading.
Voltages originating from impinging photons and the current loading on through the SNSPD are illustrated in b$5$.
The voltage trace originating from the current loading through the SNSPD is also measured with a $400$ MHz bandwidth oscilloscope for three different maximum bias currents and shown in (c).
\red{Solid line represents a single trace where the shaded area indicate electrical fluctuations.}
Trace of the current unloading voltage is similar to the current loading, but with opposite polarity.
}
\label{Fig:SetupElec}
\end{figure}

To achieve gated operation, we have designed and built the required bias control electronics. 
The circuitry is fully functional at $2.3\,\mathrm{K}$, such that it can be mounted in close proximity to the detector \red{to minimise the effect of electrical dispersion and noise pick-up from electrical connections.} 
\red{All the used electrical components are commercial off-the-shelf components that through testing are known to work at $2.3\,\mathrm{K}$.
Active elements are chosen from the OnSemi NC-7 logic family, resistors from the Vishay CH and FC series, and attenuators from the Susumu PAT series.}
Figure~\ref{Fig:SetupElec}~(a) shows a schematic of the electrical circuit and Fig.~\ref{Fig:SetupElec}~(b) shows stylised traces of the electrical signals at markers $1$~to~$5$, visualising the designed behaviour.
The electrical circuit consists of two major sections, marked as PCB~$1$ and PCB~$2$ in Fig.~\ref{Fig:SetupElec}~(a). 
The first section generates narrow pulses from an input pulse and is based on Ref.~\cite{Widhalm2018}, while the second part buffers and transform the narrow voltage pulse to the current pulse required for the operation of the SNSPD.
The circuit is triggered by an input pulse from an external function generator.
This input voltage pulse~(Fig.~\ref{Fig:SetupElec}~(a$1$)) is buffered with a Schmitt-trigger and split into two paths~(Fig.~\ref{Fig:SetupElec}~(a$2$)): the bottom path is transmitted unaltered, while the top path is inverted and delayed.
The time delay can be tuned by changing the supply voltage ($\mathrm{V_\tau}$) of the inverter.
A NAND gate recombines the two paths to generate a narrow voltage pulse with the width of the time delay imposed on the top path~(Fig.~\ref{Fig:SetupElec}~(a$3$)). 
These voltages fluctuate between ground  and the supply voltage of the PCB.
The narrow pulse is transmitted into the second part where it is buffered with an open-drain inverter and attenuated by a series of $50\mathrm\,{\Omega}$ $\mathrm{\pi}$-pad attenuators interconnected with RF-resistors. 
The attenuator is connected to the SNSPD through a serial resistor ($\mathrm{R_s}$) to transform the attenuated voltage signal into a current~(Fig.~\ref{Fig:SetupElec}~(a$4$)) while a parallel resistor ($\mathrm{R_p}$) shunts the SNSPD.
Room temperature amplifiers then amplify the voltage pulses from the SNSPD so they can be recorded by a Time to Digital Converter (TDC) or an oscilloscope~(Fig.~\ref{Fig:SetupElec}~(a$5$)).
The bias current through the SNSPD is now set by two parameters, the combination of the attenuation circuit with the serial resistor and the pull-up voltage $\mathrm{V_b}$.
Since the attenuators and serial resistor are fixed once the circuitry is mounted inside the cryostat, they set the sensitivity of the bias current.
Tuning of the bias current while the electronics are under cryogenic conditions is then performed by changing the pull-up voltage, which we henceforth refer to as the bias voltage.

Since the SNSPD has an inherent characteristic inductance~\cite{Chiles2020}, loading and unloading the current through the SNSPD causes a voltage, which is also stylistic visualised in ~(Fig.~\ref{Fig:SetupElec}~(b$5$).
A measurement of the current loading voltage is shown in Fig.~\ref{Fig:SetupElec}~(c) for three different maximum currents, namely no current~($\mathrm{I_b}=0\,\mathrm{ \mu A}$), current which maximises internal detection efficiency~($\mathrm{I_b}=19.2\,\mathrm{ \mu A}$), and an intermediate current~($\mathrm{I_b}=13.4\,\mathrm{ \mu A}$).
The traces are recorded with a $400\,\mathrm{MHz}$ bandwidth oscilloscope after the two room temperature amplifiers~($10\,\mathrm{MHz}$ to $2500\,\mathrm{MHz}$ with $30\,\mathrm{dB}$ gain each).
The trace without bias current still shows a voltage modulation which comes from parasitic leak through in the PCB.

Since the goal of an SNSPD is to detect single-photon events, four different cases are stylistically shown in Fig.~\ref{Fig:SetupElec}~(b$5$).
The cases are shown in order: no photon impinging, photon impinging while at the set bias current, photon impinging while current is loading onto the SNSPD, and photon impinging while the current is unloading from the SNSPD.
The arrival time of the photons are marked with dashed vertical lines, and the threshold voltage of the TDC is marked with a dashed horizontal line. 
This threshold voltage is important since only events where the electrical signal exceeds the threshold voltage are recorded by the TDC. 
In the event of a photon arriving when the bias current is at an appropriate set point, the voltage induced from the detection event exceeds the threshold voltage and the arrival time of the photon is recorded by the TDC.
When the photon impinges the detector while the current is loading or unloading, two effects come into play.
The first effect is that the reduced bias current causes a reduction in the voltage amplitude of the detection event.
The second effect is that the voltage caused by the current loading is still non-zero, causing an offset to the voltage from the detection event.
This combination causes an irregular behaviour of the maximum voltage of a detection event depending on the detection time of the photon by the SNSPD, where not every arrival time on the rising or falling edge of the current pulse necessarily causes an event to be registered by the TDC.

Reducing the loading and unloading time of the current minimises the times at which photons may be absorbed but not registered by the TDC.  
This reduction also allows narrower gate windows and faster repetition of the gating.
An analytical model of the current loading is extracted from a simplified electrical circuit by replacing the attenuation circuit by a voltage divider, and the bias voltage and generated pulses with a function generator\red{, as shown by the model in Fig.~\ref{Fig:ElecModel}}.
Justification and additional information for this model is found in Supplementary material $1$.

\begin{figure}[ht!]
\centering\includegraphics{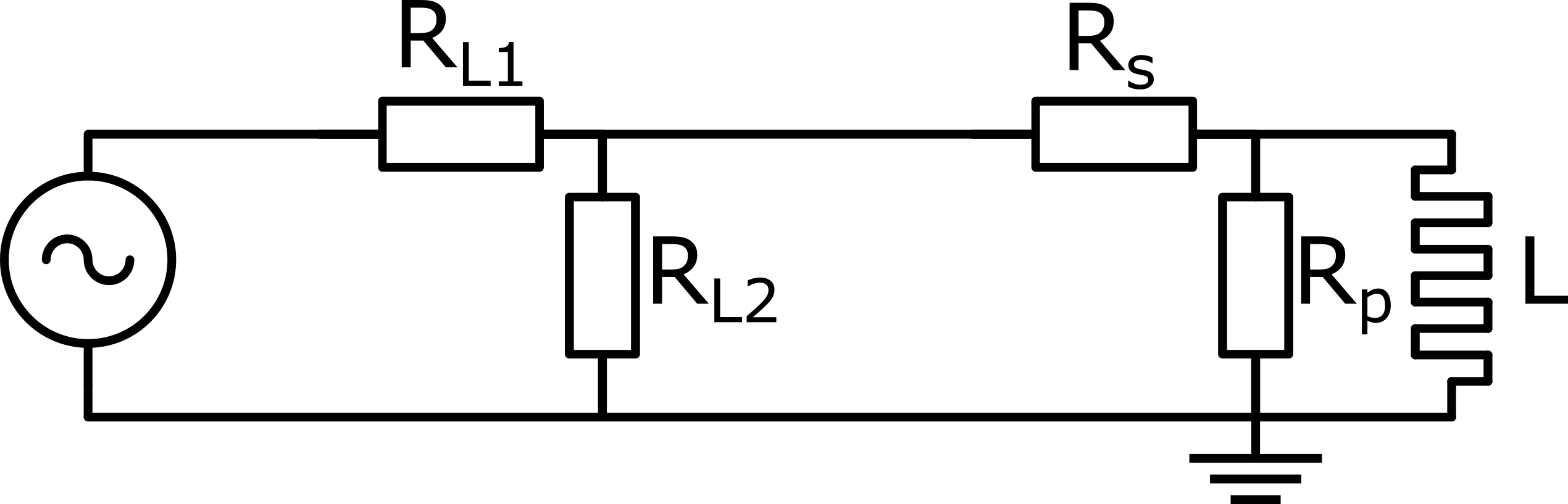}
\caption{Simplified circuit to model the current loading and unloading through the SNSPD.
The attenuator is modelled with a voltage divider constructed from $\mathrm{R_{L1}}$ and $\mathrm{R_{L2}}$, and the bias current generation is replaced with a function generator.
This simplification is valid when the attenuator can be considered a constant impedance for the SNSPD combined with the parallel $\mathrm{R_p}$ and serial $\mathrm{R_s}$ resistor.
}
\label{Fig:ElecModel}
\end{figure}

The current loading follows an inverted exponential decay with time constant $\tau_{\mathrm{rm}}$, where this time constant is extracted as
\begin{equation}
\tau_{\mathrm{rm}} =
 \frac{
 \mathrm{L}\left(
 \mathrm{R_{s}}\left(2\mathrm{R_{L1}+R_{L2}}\right)+\mathrm{R_{p}\left(R_{L1}+R_{L2}\right)}
 \right)}
 {\mathrm{R_p R_s}\left(2\mathrm{R_{L1}+R_{L2}}\right)}~,
 \label{EQ:LoadTime}
\end{equation}
with $\mathrm{R_{L1}}$ and $\mathrm{R_{L2}}$ as the substitute resistors for the attenuator, $\mathrm{R_p}$ and $\mathrm{R_s}$ are respectively the parallel and serial resistors, and $\mathrm{L}$ as the SNSPD inductance.
This shows that the current loading time is directly affected by the inductance of the SNSPD, but that it can be modified with selecting a proper combination of $\mathrm{R_p}$ and $\mathrm{R_s}$.

Selecting the proper components for the gating circuitry is a trade-off between the desired detection and gating properties.
Reduction of the current loading times require an increased $\mathrm{R_s}$ and $\mathrm{R_p}$, however, increasing $\mathrm{R_p}$ can result in a loss of self-resetting of the SNSPD~\cite{Yang2007}.
Since our aim is to compare the characteristics of the SNSPD in gated and DC operation, we selected $\mathrm{R_p}=330\,\mathrm{\Omega}$ and $\mathrm{R_p}=1000\,\mathrm{\Omega}$.
The attenuator is designed to yield $\mathrm{I_b}/\mathrm{V_b}=20.0\,\mathrm{ \mu A /V}$ since previous tests on the SNSPD indicated operating currents in the range from $10$ to $20$ $\mathrm{\mu A}$.
The measured dependency between the bias current and bias voltage at $2.3\,\mathrm{K}$ is $\mathrm{I_b}/\mathrm{V_b}=(19.18\pm0.07)\,\mathrm{ \mu A /V}$.
\red{The induced heatload from the PCBs is measured between $1\,\mathrm{mW}$ to $3\,\mathrm{mW}$, and shows strong dependence on the chosen attenuation circuit.}

\section{Characterising gated operation of an SNSPD} \label{CharGateMode}
\begin{figure}[ht!]
\centering\includegraphics{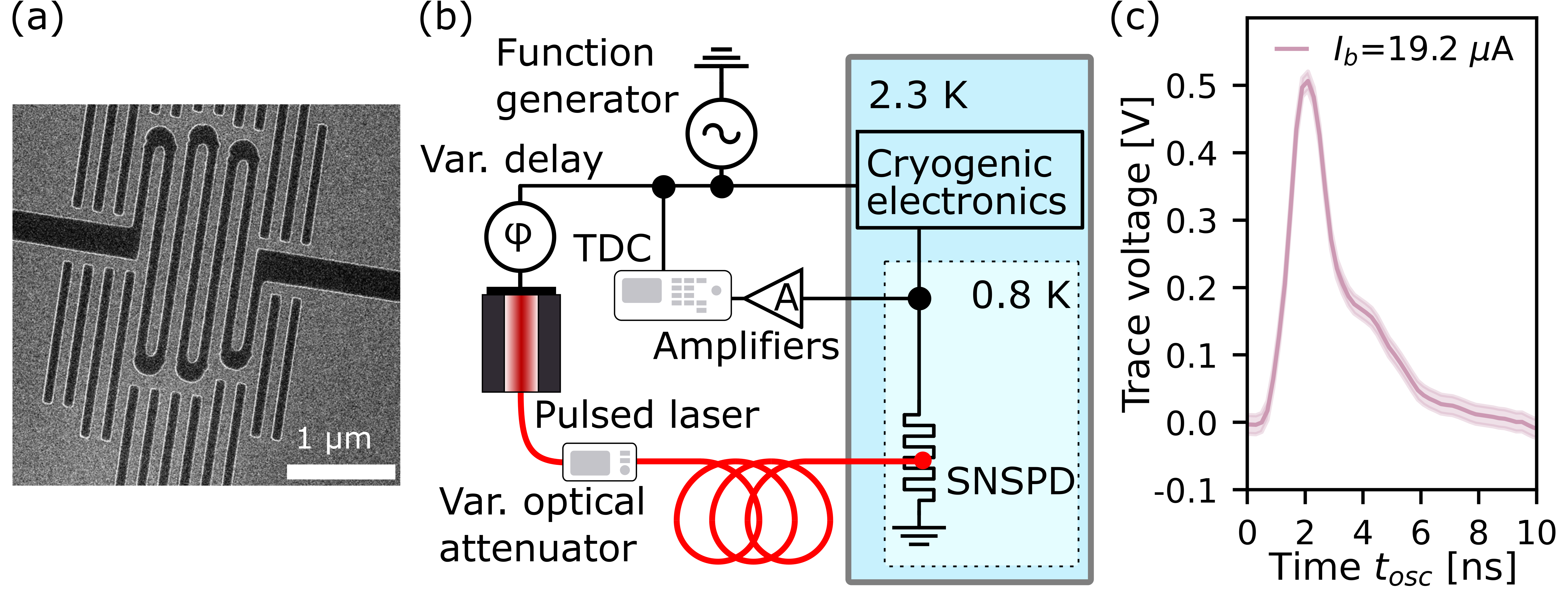}
\caption{
(a) SEM image of the SNSPD used in this experiment.
(b) Schematic of the optical setup used to characterise the SNSPD in gated operation mode.
A function generator triggers the cryogenic electronics (described in Fig.~\ref{Fig:SetupElec}~(a)), as well as a pulsed laser via a variable delay.
The pulsed laser is attenuated through a variable optical attenuator to achieve detected optical powers in the single-photon regime.
The electrical pulses from the SNSPD are amplified and the timestamps are recorded with a TDC, which also records the timestamps of the trigger pulses.
(c) Trace of a photon detection event recorded with a $400$ MHz bandwidth oscilloscope. 
\red{Solid line represents a single trace where the shaded area indicate electrical fluctuations.}
}
\label{Fig:Setup_Optic}
\end{figure}

The SNSPD we use is patterned from a $9\,\mathrm{nm}$ thick magnetron co-sputtered NbTiN film with a strip width of $100\,\mathrm{nm}$.
The detector surface area has a diameter of $2\,\mathrm{\mu m}$ and fill-factor of $50\%$.
\red{The used film thickness originates from previous work~\cite{Zichi2019} indicating lower kinetic inductance and increased optical absorption.}
An SEM image of the SNSPD can be seen in Fig.~\ref{Fig:Setup_Optic}~(a).
To characterise the performance of gating this device, it is mounted on the cold stage of a closed cycle sorption cryostat at $0.8\,\mathrm{K}$.
The electronics are mounted on a stage at 2.3K since this offers a higher cooling power.
\red{A function generator triggers the cryogenic electronics, the TDC ($10$~ps resolution), and via a variable time delay the pulsed illumination laser (laser diode, $<50ps$~FWHM).}
The pulsed laser is attenuated by a variable optical attenuator, coupled into the cryostat, and coupled to the SNSPD through flood illumination.
The voltage over the SNSPD is amplified by two room temperature amplifiers~($10\,\mathrm{MHz}$ to $2500\,\mathrm{MHz}$ with $30\,\mathrm{dB}$ gain each) to record the detection times with the TDC.
This is schematically shown in Fig.~\ref{Fig:Setup_Optic}~(b).

Unless otherwise specified, all characterisation experiments are \red{performed} with a system repetition rate of $1.0\,\mathrm{MHz}$ and a trigger pulse width of $500\,\mathrm{ns}$.
This architecture yields gating windows narrower than the trigger pulse width, although longer gates, and indeed DC operation, are possible with different repetition rates. 
We ensure that the trigger pulse has a maximum duty cycle of $50\%$ to avoid unwanted artefacts when recombining the two pulses at the NAND gate (stage $3$ in Fig.~\ref{Fig:SetupElec}~(b)).
The optical input power into the cryostat is set to \red{and measured as } $\mathrm{E_p}=(7.1\pm0.1)\cdot10^{-13}\,\mathrm{\frac{J}{Pulse}}$ at a wavelength of $1550\,\mathrm{nm}$.
\red{Losses due to the flood illumination of the SNSPD are modelled based on relative position and angle between the fibre and SNSPD, measured during mounting inside the cryostat.
These models predict a coupling loss of $(85\pm3)\,\mathrm{dB}$, resulting in a}
mean photon number of $\left(18^{+18}_{-9}\right)\cdot10^{-3}\,\mathrm{\frac{photons}{pulse}}$ incident on the detector.

A voltage trace of a photon detection event is recorded with a $400\,\mathrm{MHz}$ bandwidth oscilloscope is shown in Fig.~\ref{Fig:Setup_Optic}~(c), and has a peak around $500\,\mathrm{mV}$ after the amplifiers.
This shows that a single-photon detection event reaches higher amplitudes than the current loading voltage ($320\,\mathrm{mV}$, Fig.~\ref{Fig:SetupElec}~(c)) indicating there is a range of threshold voltages for the TDC on which only detection events are recorded.
The ratio between the voltage of a single-photon detection event and the current loading voltage depends on the inductance of the SNSPD and the control electronics around the SNSPD.

\subsection{Bias current dependant \red{count rate}}
\begin{figure}[ht!]
\centering
\includegraphics{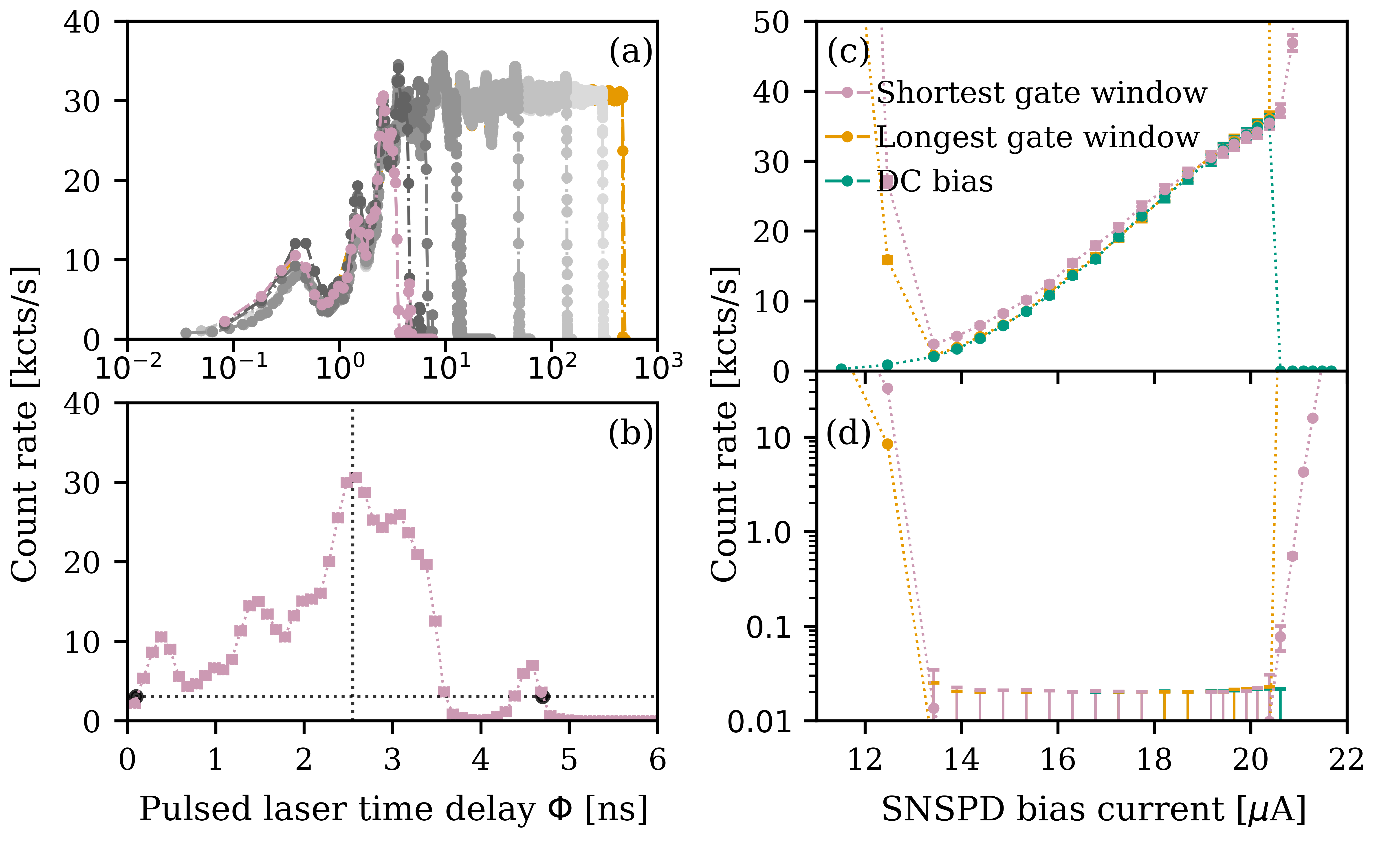}
\caption{Characterisation of the temporal profile of different gate window lengths and comparison between DC-operation and gated operation.
(a) Temporal profile of different gate window lengths with highlighted the longest possible gate window in our operation ($500\,\mathrm{ns}$, yellow) and the shortest possible gate window in our operation ($5.0\,\mathrm{ns}$, pink).
(b) The temporal profile of the narrowest gate window with linear time axis.
The rugged shape is a combination of multiple effects (see text).
The dotted vertical line indicate the current loading time ($<2.4\,\mathrm{ns}$) and thus the rise time, and the dotted horizontal line indicate the total pulse length ($<5.0\,\mathrm{ns}$).
\red{(c)}
Bias current dependent \red{count rate} for three different operation modes: Shortest Gate window (SG), Longest Gate window (LG), and DC operation.
\red{(d) Background count rates, measured without illumination.}
Threshold voltage of the TDC is adjusted to the set bias current.
}
\label{Fig:CharPhasePlat}
\end{figure}

Comparing the \red{count rate} of the SNSPD between gated operation and biasing with a direct bias current (DC-operation) require a combined characterisation of the tuning of the gate window and optimisation of the bias current.
First we investigate the tuning range and profile of the gate window by fixing the delay voltage $\mathrm{V_{\tau}}$, scanning the time delay of the laser trigger $\mathrm{\phi}$, and recording the detected count rate from the SNSPD with a TDC.
Figure~\ref{Fig:CharPhasePlat}~(a) shows the temporal profiles for different gate windows with a logarithmic time axis and a bias current of $\mathrm{I_b}=(19.18\pm0.07)\,\mathrm{ \mu A}$.
The pink line indicates the Shortest Gate window (SG) we can generate while still reaching maximum count rate, the yellow line marks the Longest Gate window (LG) we can generate, and the shades of grey show that we can tune the gate window in-between these limits.
The longest gate we generated is limited by the setup parameters of a repetition rate of $1.0\,\mathrm{MHz}$ and $50\%$ duty cycle, and could be increased by decreasing the repetition rate.
The shortest gate window is re-plotted in Fig.~\ref{Fig:CharPhasePlat}~(b) with a linear time axis and markers at relevant points.
The profile of the shortest gate does not follow a smooth inverted exponential decay due to various reasons.
First, as previously discussed in section \ref{GatingCircuit}, there exists a base line offset due to the current loading voltage.
Secondly, the \red{count rate} of the SNSPD is reduced at lower bias currents, thus while the current is loading, the \red{count rate} is increasing towards its maximum.
%
A third reason comes from electrical resonances causing current fluctuations, which manifest as a time dependant \red{count rate} if the \red{count rate} of the SNSPD does not saturate with the bias current.
These three reasons affect either the \red{count rate} directly or the maximum voltage of a detection event, which vary the recorded count rate since the TDC has a fixed threshold voltage.
The interplay of these effects may explain the rugged shape visible in the narrowest gating window from Fig.~\ref{Fig:CharPhasePlat}~(b), however, from the narrowest pulse it can be seen that the current is fully loaded through the SNSPD in $\left(2.4\pm0.1\right)\,\mathrm{ns}$ as marked by the vertical dotted line.
It can also be derived that the $10\%$ to $10\%$ pulse width is less than $5.0\,\mathrm{ns}$, as marked by the dashed horizontal line in Fig.~\ref{Fig:CharPhasePlat}~(b).
The pulse width is measured to the after pulse between $4.2\,\mathrm{ns}$ and $4.9\,\mathrm{ns}$ since the drop in count rate between $3.5\,\mathrm{ns}$ and $4.2\,\mathrm{ns}$ is caused by the reduced voltage background from the current unloading, thus hiding photon detection events from the TDC.
Using this data and the model from Eq.~\ref{EQ:LoadTime}, we extract a maximum inductance of the SNSPD of $260\,\mathrm{nH}$.

Secondly, we compare the bias current dependant \red{count rate} between the gated and DC operations with three different settings: DC biasing of the SNSPD, $500\,\mathrm{ns}$ gate window (LG) with an optical pulse at $200\,\mathrm{ns}$, and a $5.0\,\mathrm{ns}$ gate window (SG) with the optical pulse at $2.4\,\mathrm{ns}$.
A DC bias current is created by supplying a bias voltage $V_b$ into the attenuator without turning on the pulsing section of the gating circuit.
The bias current is set by tuning the bias voltage, and the trigger threshold of the TDC is adjusted accordingly since the amplitude of detection events reduces with a lower bias current.
Figure~\ref{Fig:CharPhasePlat}~(c) shows the total detection count rate for the three different cases over a bias current range of \red{$11\,\mathrm{\mu A}$ to $22\,\mathrm{\mu A}$} when the SNSPD is illuminated with the pulsed laser \red{($<50\,\mathrm{ps}$ pulse width,} optical power as discussed in section~\ref{CharGateMode}).
From a bias current between $\mathrm{I_b}=(13.4\pm0.1)\,\mathrm{\mu A}$ and $\mathrm{I_b}=(20.5\pm0.1)\,\mathrm{\mu A}$ all three operation modes show close agreement in detected count rates.
Beyond $\mathrm{I_b}=(20.5\pm0.1)\,\mathrm{\mu A}$ the DC bias mode drops to no detection events which indicates that the SNSPD has reached a fully latched state.

\red{Reaching the latched state is a dynamic process in which the SNSPD settles into an electrothermally stable and photon-insensitive regime, which is reached when the bias current is larger than the latching current ($I_L<I_b$) or when the bias current is larger than the critical current ($I_c<I_b$).
However, when the latching current of the SNSPD is larger than the critical current, a train of electrical self-oscillations occur if the bias current is between the critical current and latching current ($I_c<I_b<I_L$)~\cite{Kerman2009,Kerman2013}.
The self-sustainability of these oscillations depends on the bias circuitry, and the self-oscillation frequency depends on the bias current.
In a non self-sustainable configuration, the DC operation latches the detector while gating resets the self oscillations and can thus maintain them.
This results in an increased background count rate for gated operation while DC operation latches when the bias current exceeds the critical current.
}

We measure the background counts under the same operating conditions but without illumination from a laser.
This allows us to measure illumination independent effects such as \red{dark counts, current loading voltages, and electrothermal self-oscillations}.
\red{This bias current dependant background count rate is shown in  Fig.~\ref{Fig:CharPhasePlat}~(d),}
from which it can be seen that the increase in count rate in gated operation mode with $\mathrm{I_b}>20.5\,\mathrm{\mu A}$ comes from an increase in background counts \red{indicating a critical current of $\mathrm{I_c}=20.5\,\mathrm{\mu A}$ in all three cases}.
A bias current $\mathrm{I_b}<13.4\,\mathrm{\mu A}$ also yields an increase in the background counts in gated operation, which is induced via the electrical leak through in the PCB as discussed in Fig.~\ref{Fig:SetupElec}~(c).
The amplitude of these leak-through pulses is measured to be independent of the bias current, which is expected since the bias current is not the source of this noise.
The voltage amplitude of the detection events scales directly with the bias current, related to each other through the parallel resistor $\mathrm{R_p}$.
Smaller bias currents thus results in lower amplitudes of a photon detection event. 
When this amplitude approaches the amplitude of the current loading voltage plus the leak-through, it becomes impossible to discriminate between them using the TDC, since no threshold voltage can be set to trigger only on a photon detection event.
\red{This background count rate will thus reach the repetition rate for lower bias currents.}
Reduction of the leak-through can be implemented in future designs by careful analysis of the origin of the parasitic coupling, and then minimising this source.

\red{The increase in background counts in Fig.~\ref{Fig:CharPhasePlat}~(d) originates from self-oscillations of the SNSPD, indicating that the bias current and latching current are larger than the critical current while the bias current is still smaller than the latching current~\cite{Kerman2013}.
Since the critical current is reached before the detector saturates in the bias current dependant count rate, maximum internal detection efficiency is not reached.}
This means that the SNSPD is prone to temporal detection efficiency changes based on bias current fluctuations.
This can be overcome by improving fabrication techniques such that the internal detection efficiency does saturate.


\subsection{Timing jitter}
\begin{figure}[b!]
\centering
\includegraphics{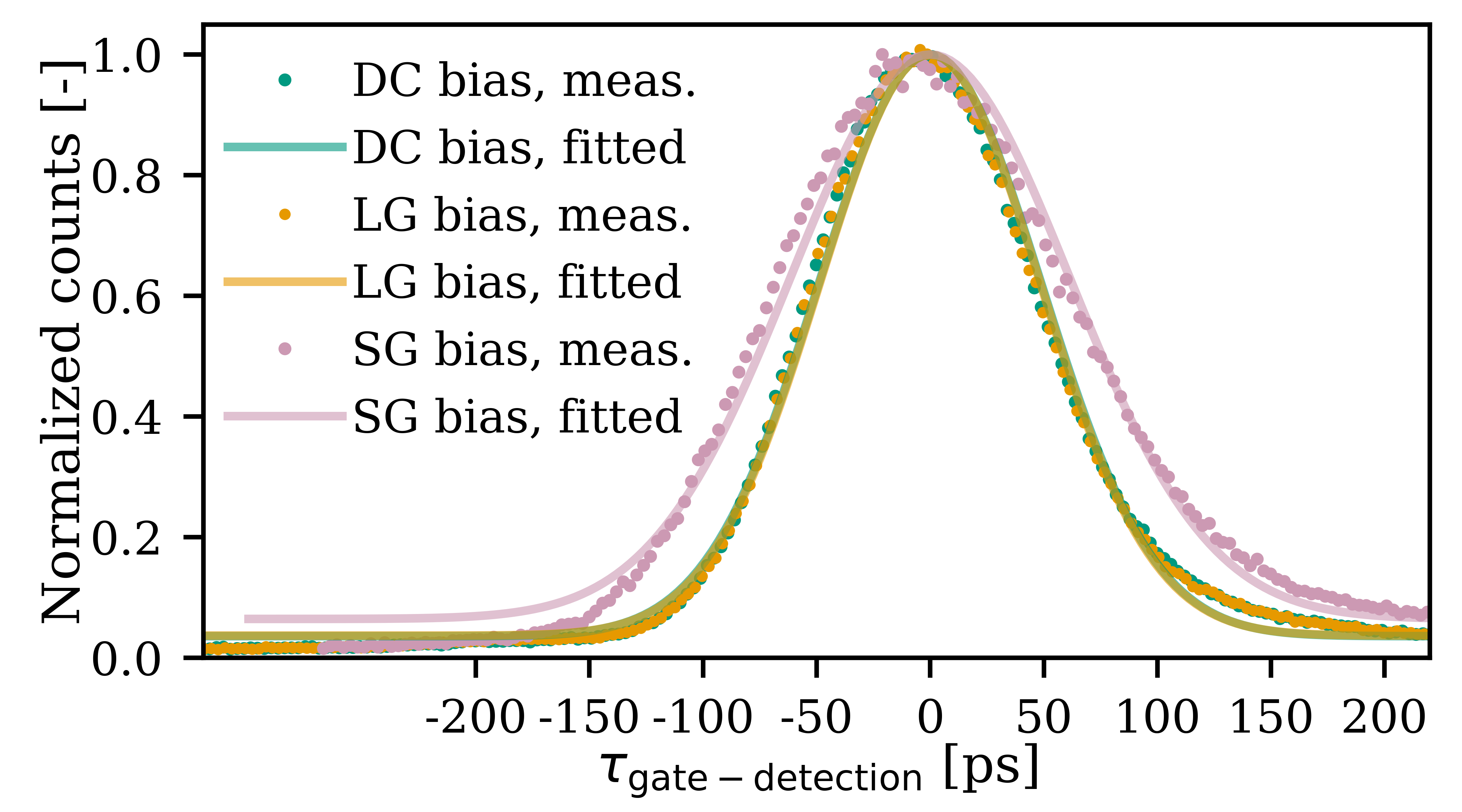}
\caption{Measured timing jitter of photon detection events from the SNSPD for the three different operation modes (DC operation, Longest Gate window (LG), and Shortest Gate window (SG)).
Dots indicate the the measured counts and the lines mark the Gaussian fits.
The Shortest gate window increased the timing jitter by approximately $30\,\mathrm{ps}$.
}
\label{Fig:CharJitter}
\end{figure}
One of the advantages of SNSPDs is their low timing jitter \cite{korzh2020}, which has to be preserved when gating the SNSPD.
The total system timing jitter is derived as the fluctuation in the measured time difference between the gating trigger and the electrical response of the SNSPD, when all parameters are kept constant.
Thus, the total system jitter include all timing jitter effects in the setup such as the electrical trigger signal, optical pulse generation, amplifiers, and timetagger.
The Full Width Half Maximum (FWHM) of the system jitter is extracted through the standard deviation from a Gaussian fitting function.
The measured time fluctuation and the Gaussian fits of all three operation modes are shown in Fig.~\ref{Fig:CharJitter} where the dots are the measured histograms and the lines the Gaussian fits.

The timing jitter in DC and long gate operation mode are essentially identical and extracted as $\mathrm{\tau_{DC}}=(115.23\pm0.15)\,\mathrm{ps}$ and $\mathrm{\tau_{LG}}=(114.30\pm0.13)\,\mathrm{ps}$.
Narrowing the gate window induced an increase in the jitter of $(29.1\pm0.6)\,\mathrm{ps}$ to $\mathrm{\tau_{SG}}=(144.3\pm0.4)\,\mathrm{ps}$.
The proximity between the DC and long gate operation show that the \red{increase} in timing jitter in the short gate is not intrinsic to the gating circuit \textit{per se} \red{as long as the temporal profile of the light is completely in the constant on time of the gating window.
}
Since the bias current has no constant region in the \red{shortest gating} mode, the amplitude of the detection signal depends on when the photon was absorbed by the SNSPD.
The threshold voltage for the TDC is however at a fixed voltage and the TDC therefore does not always trigger at the same relative height in the amplitude.
This fluctuation translates in a changed detection time.
This chain of processes therefore amplifies any timing jitter between the photon arrival time and the gate window when the SNSPD is operated in \red{the shortest} gating mode.
This may be overcome with a different measurement apparatus such as a fast real-time oscilloscope and flank triggering \red{such that the timing change due to a fixed threshold voltage can be negated}.
Furthermore, using cryogenic amplifiers can reduce the \red{electronic jitter, and thus the} system jitter.

\subsection{Autocorrelation}
\begin{figure}[b!]
\centering\includegraphics{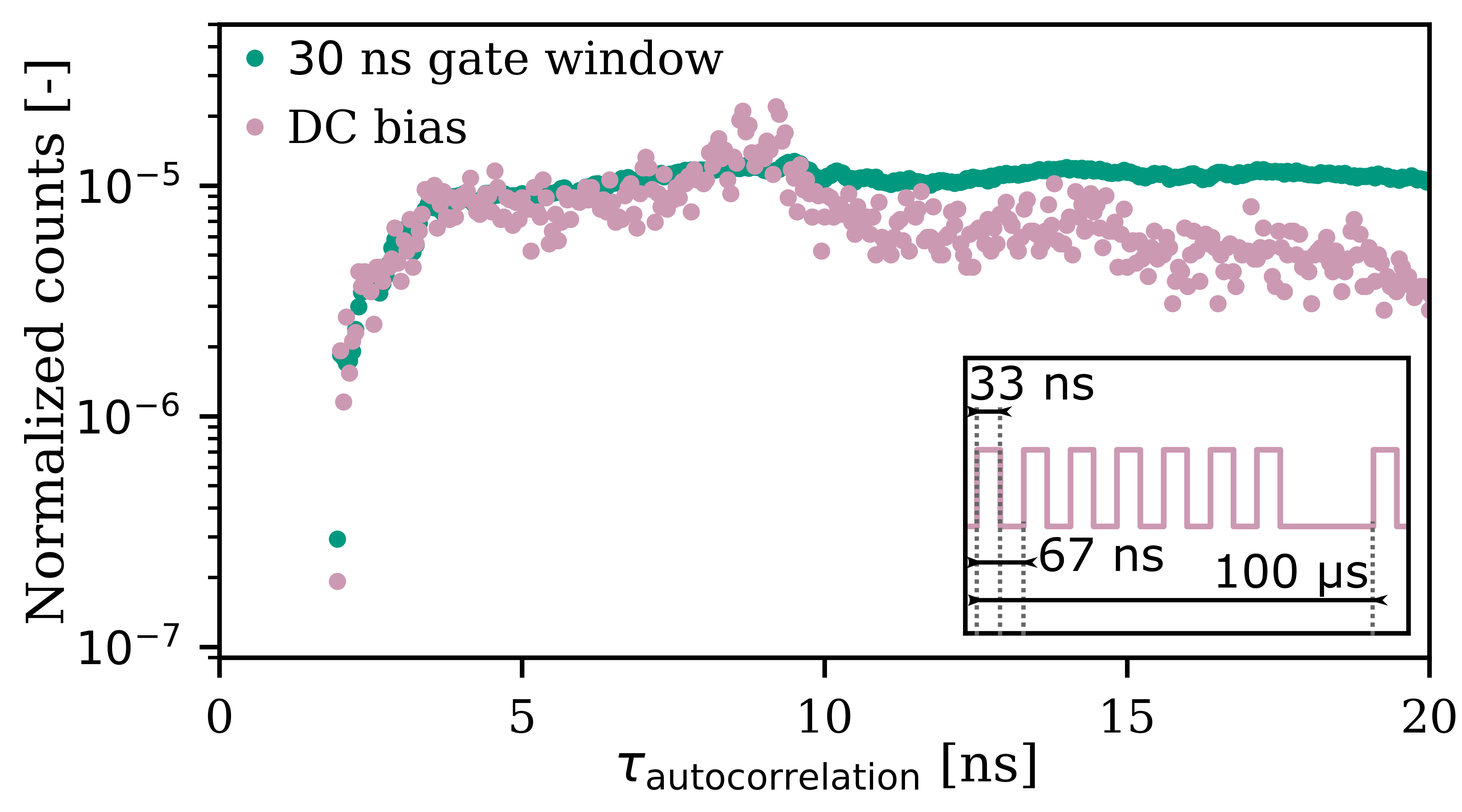}
\caption{Auto-correlation of the SNSPD in DC and gated operation\red{, normalised to $\tau=0$ to compare both operation modes}.
The SNSPD has a $50\%$ recovery time of $(3.0\pm0.1)\,\mathrm{ns}$, which is independent of the operation mode.
Inset shows the trigger sequence used in gated mode for the autocorrelation.
}
\label{Fig:CharAutocorr}
\end{figure}
The recovery time of a single-photon detector is measured with an autocorrelation on the photon detection events~\cite{Autebert2020}.
The setting for the autocorrelation measurements are different than for the previous measurements.
The SNSPD is illuminated with \red{continuous wave (CW)} light at a wavelength of $1550\,\mathrm{nm}$ with an optical power of $2.4\,\mathrm{\mu W}$ at the input of the cryostat resulting in an expected photon detection rate of $64\,\mathrm{kcts/s}$.
The gating circuit is triggered in burst operation with $7$ pulses at $15\,\mathrm{MHz}$ every $100\,\mathrm{\mu s}$ and the gating window is set to a duty cycle of $50\%$ (see inset Fig.~\ref{Fig:CharAutocorr}), which is compared to a DC biased SNSPD.
A zoom in on the autocorrelation of both operation modes over a time window of $20\,\mathrm{ns}$ is shown in Fig.~\ref{Fig:CharAutocorr} from which a $50\%$ SNSPD recovery time of $(3.0\pm0.1)\,\mathrm{ns}$ can be extracted.
Further, the overlap between the two modes in the sub $8\,\mathrm{ns}$ shows that the gating neither improves nor degrades the recovery time.
After $10\,\mathrm{ns}$ the gated autocorrelation decreases, since the autocorrelation of a square pulse is triangular.
An increased correlation in the gated mode at $(8.8\pm0.8)\,\mathrm{ns}$ is observed, the origin of which can be parasitic current resonance.

\subsection{Characterisation summary}
The characterisation of the SNSPD and the gate window show that our electrical circuit allows the tuning of a gate window of an SNSPD down to $5.0\,\mathrm{ns}$ with a rise time of $(2.4\pm0.1)\,\mathrm{ns}$.
From this rise time and the simulation we can calculate that we have an expected inductance of $\mathrm{L_{SNSPD}}<260\,\mathrm{nH}$.
The gating of our SNSPD does not change the \red{count rate} compared to a DC biased SNSPD in a typical bias regime.
The total system timing jitter in DC operation mode is $\mathrm{\tau_{DC}}=(115.23\pm0.15)\,\mathrm{ps}$ and with the shortest gate window $\mathrm{\tau_{SG}}=(144.3\pm0.4)\,\mathrm{ps}$.
The increased of $(29.1\pm0.6)\,\mathrm{ps}$ can be explained with the proximity between the photon arrival time and the varying current from the current loading and unloading.
The recovery time of the detector is not changed by the gating, however, the recovery time is longer than the current loading time, which is promising for temporal filtering of pump light and high dynamic range photon counting.

\section{Exploiting gated SNSPDs}
Two proof of principle experiments are performed to demonstrate the benefits of deterministic gating of an SNSPD.
We first demonstrate high dynamic range photon counting, where a gated SNSPD is used to measure a higher mean photon number than the count rate limit of the detector in DC biased mode. 
Secondly we demonstrate the temporal filtering of a pump pulse $2.8\,\mathrm{ns}$ before a signal pulse.

\subsection{High dynamic range photon counting}
Our approach to high dynamic range photon counting of continuous wave light is based on relating the distribution of photon arrival times to mean photon number. 
We use the principle that the waiting time to measure the first photon depends on the photon flux of the optical signal, as well as the underlying photon statistics of the source. 
This waiting time can be measured with a gated SNSPD, where the start time is the start of the gate, and the stop time is the arrival of the first photon.
This measurement probes the Probability Density Function (PDF) $p_g\left( t \right)$ of the arrival time of the first photon after the start of the gate window \red{and is dependant on the inter-arrival time between photons $p\left( t \right)$ and the probability distribution of the waiting time for the gate window to start after the last arrived photon $p_w\left( t \right)$}.

\red{Deriving the inter-arrival time PDF between photons can be done by calculating the probability that the second} photon arrives in a time window from $t=0$ to $t=T$ \red{after the first photon, which} is calculated by integrating the PDF over this window $P\left( T \right) = \int_{0}^{T}p\left( t \right) \mathrm{d} t$.
This definition can be written in differential form $p\left(t\right)=\frac{\mathrm{d} }{\mathrm{d} t}P\left(T=t\right)$.
Using these definitions, the probability that the \red{second} photon does \emph{not} arrive before time $T$ can thus be calculated by $P^\prime\left( T \right) = 1-P\left(T\right) = 1- \int_{0}^{T} p\left( t\right) \mathrm{d}t = \int_{T}^{\infty} p\left( t\right) \mathrm{d}t$, which is the right hand side cumulative probability. 
In differential form, we can write $p\left(t\right)=\frac{\mathrm{d}}{\mathrm{d} t} P\left(T=t\right)=\frac{\mathrm{d}}{\mathrm{d} t}\left(1-P^\prime\left(t\right)\right)=-\frac{\mathrm{d}}{\mathrm{d} t}P^\prime\left(t\right)$.

Crucially, the probability that a photon does not arrive in a time $T$ is the same as the probability of measuring the vacuum state given the underlying photon statistics $\mathcal{P}\left(n\right)$ of the impinging light, {i.e.} $P^\prime\left(T\right)=\mathcal{P}\left(n=0\right)$. 
Thus we can relate the \red{photon inter-arrival} time PDF, to the photon statistics as
\begin{equation}\label{eqn:pdf}
p\left(t\right)=-\frac{\mathrm{d}}{\mathrm{d} t}\mathcal{P}\left(0\right)~.
\end{equation}

The photon probability distribution $\mathcal{P}\left(n\right)$ is a function of the mean photon number $\bar{n}$ given by $\bar{n}=\Phi \cdot t$, where $\Phi$ is the  photon flux over a measurement time $t$. However the nature of this function depends on the underlying source of light. 
For coherent and thermal light, the distributions are $\mathcal{P}_\mathrm{coh}=\frac{\bar{n}^n}{n!}e^{-\bar{n}}$ and  $\mathcal{P}_\mathrm{therm}=\frac{1}{\bar{n}+1} \left( \frac{\bar{n}} {\bar{n}+1}\right)^n  $ respectively. 
Therefore we arrive at two forms of Eqn.~\ref{eqn:pdf}, namely
\begin{align}
    p_\mathrm{coh}\left(t\right)&=-\frac{\mathrm{d}}{\mathrm{d} t}
    e^{-\Phi t}=\Phi e^{-\Phi t}~,\\
    p_\mathrm{therm}\left(t\right)&=-\frac{\mathrm{d}}{\mathrm{d} t}
    \frac{1}{\Phi t+1}= \frac{\Phi}{\left(\Phi t+1\right)^2}~.
\end{align}

\begin{figure}[b!]
\centering\includegraphics{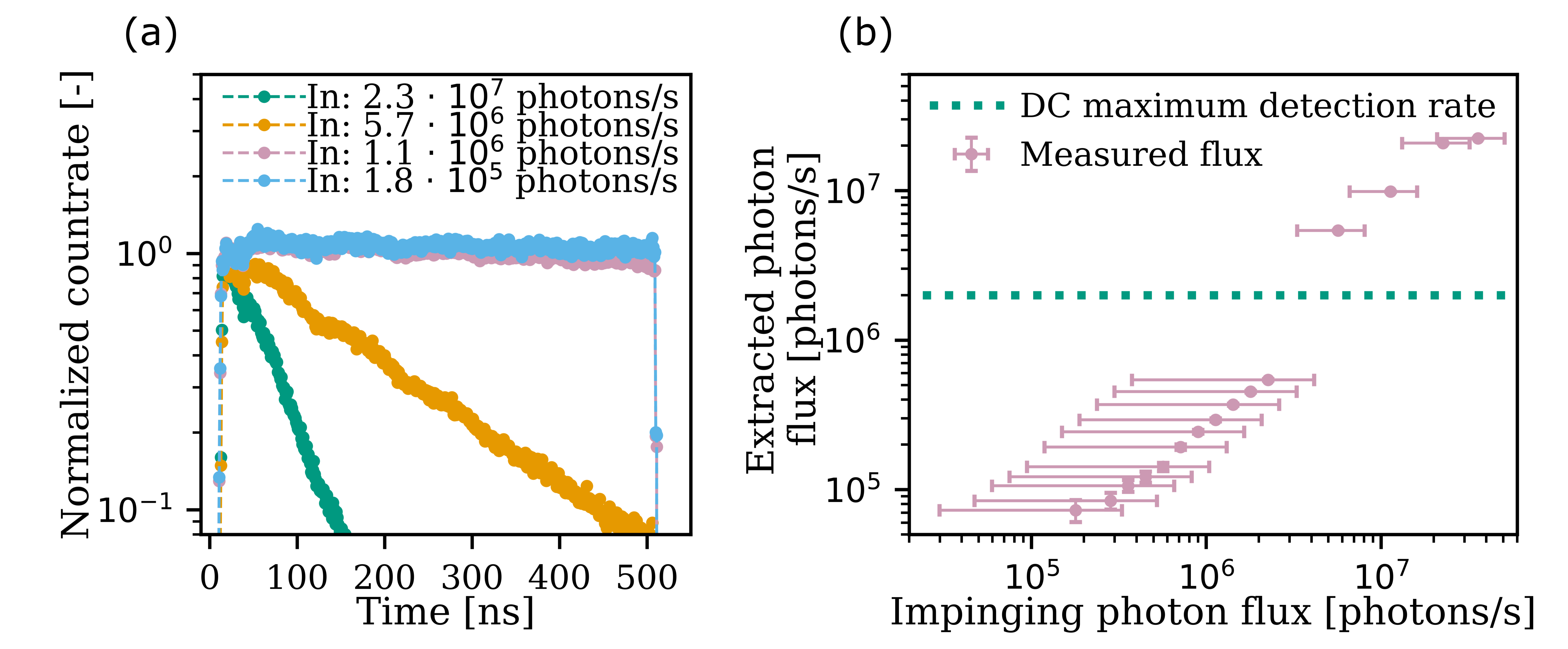}
\caption{High dynamic range photon count rate extraction with coherent light incident on the SNSPD.
(a) Normalised histogram for the first photon arrival time measurement for four different input photon fluxes.
These traces show different exponential decay constants for the different input powers.
The cutoff at $500\,\mathrm{ns}$ appears due to the limited gate window.
(b) Extracted photon fluxes for all measured input photon fluxes.
The horizontal line indicates the maximum count rate the SNSPD in this configuration could achieve with DC biasing.
The extracted photon count rate in gated operation was at least a factor $11.2\pm0.1$ higher than in DC operation.
Higher photon count rates were not achieved since the flood illumination attenuation forced optical input powers into the cryostat in the order of the maximum heat load of the cryostat.
}
\label{Fig:AppDynamic}
\end{figure}

Our aim is to infer the photon flux $\Phi$ from the first photon arrival time distributions $p\left(t\right)$ \red{in a gating experiment, for which the inter-arrival times have to be converted to a gated first photon arrival time PDF.
This conversion is made by averaging over the inter-arrival time $p\left(t\right)$ and the PDF $p_w\left(t\right)$ that the gate starts time $t_g$ after the previous arrived photon, which is given by
\begin{align}
    p_g\left( t\right) = \frac {\int_0^{\mathrm{t_r}}p_{d}(t_g)p_{w}(t_g)p(t_g+t)dt_g} {\int_0^{\mathrm{t_r}}p_{d}(t_g)p_{w}(t_g)\mathcal{P}\left(0\right)dt_g}~,
\end{align}
where $p_d$ is the temporal response of the detector and $t_r$ the maximum waiting time, thus the repetition rate in a pulsed experiment.
When the SNSPD is gated at a fixed repetition rate and the photons have a random arrival time before the start of the gating, $p_{w}$ becomes a block function with the width of the repetition rate.
Assuming the random arrival times of the photons, the measured PDF for coherent light is now transformed into a truncated exponential decay with the same decay constant during the time window where the detector shows a constant response.
This constant response can be assumed after a certain rise time (ref.~\cite{Autebert2020} or see Fig.~\ref{Fig:CharPhasePlat}~(a)), and the following shape gives us thus the necessary information to extract the photon flux.
}

Taking photons from a coherent state as an example, the gradient of the exponential decay can be fitted to return $\Phi$ directly. 
We performed this experiment using the gated detector and a continuous wave laser as a light source, and thus only extract the photon count rate for a coherent light source.

To perform this experiment, the SNSPD is biased at $\mathrm{I_b}=\left(19.18\pm0.07\right)\,\mathrm{ \mu A}$ with a repetition rate of $100\,\mathrm{kHz}$ and a gate window of $500\,\mathrm{ns}$.
A histogram of the arrival time of the first photon compared to the gate trigger is measured, an exponential decay is fitted, and the photon flux is extracted from the decay parameter. The normalised histogram for four input photon fluxes are shown in Fig.~\ref{Fig:AppDynamic}~(a), showing the clear exponential decay behaviour.
This extracted photon flux is plotted in Fig.~\ref{Fig:AppDynamic}~(b) relative to the expected detected photon flux based on optical input power and flood illumination coupling loss.
The uncertainty on the input photon counts comprise of the input flood illumination coupling loss and polarisation sensitivity of the SNSPD.
A dashed horizontal line indicates the measured upper bound of the maximum achievable count rate ($2.0\,\mathrm{M\,cts/s}$) with the SNSPD operated in DC mode at the same bias current.
Lower photon count rates randomly caused latching, but from $2.0\,\mathrm{M\,cts/s}$ the SNSPD always latched.
The waiting time measurement allowed the extraction of a photon flux up to $\left(22.4\pm0.2\right)\,\mathrm{M\,cts/s}$, which is at least a factor of $11.2\pm0.1$ more than what could be measured in DC mode.
Higher impinging photon fluxes could not be achieved in the current configuration since the flood illumination losses limited the cooling capability of the cryostat.

\subsection{Temporal pulse filtering}

\begin{figure}[b!]
\centering\includegraphics{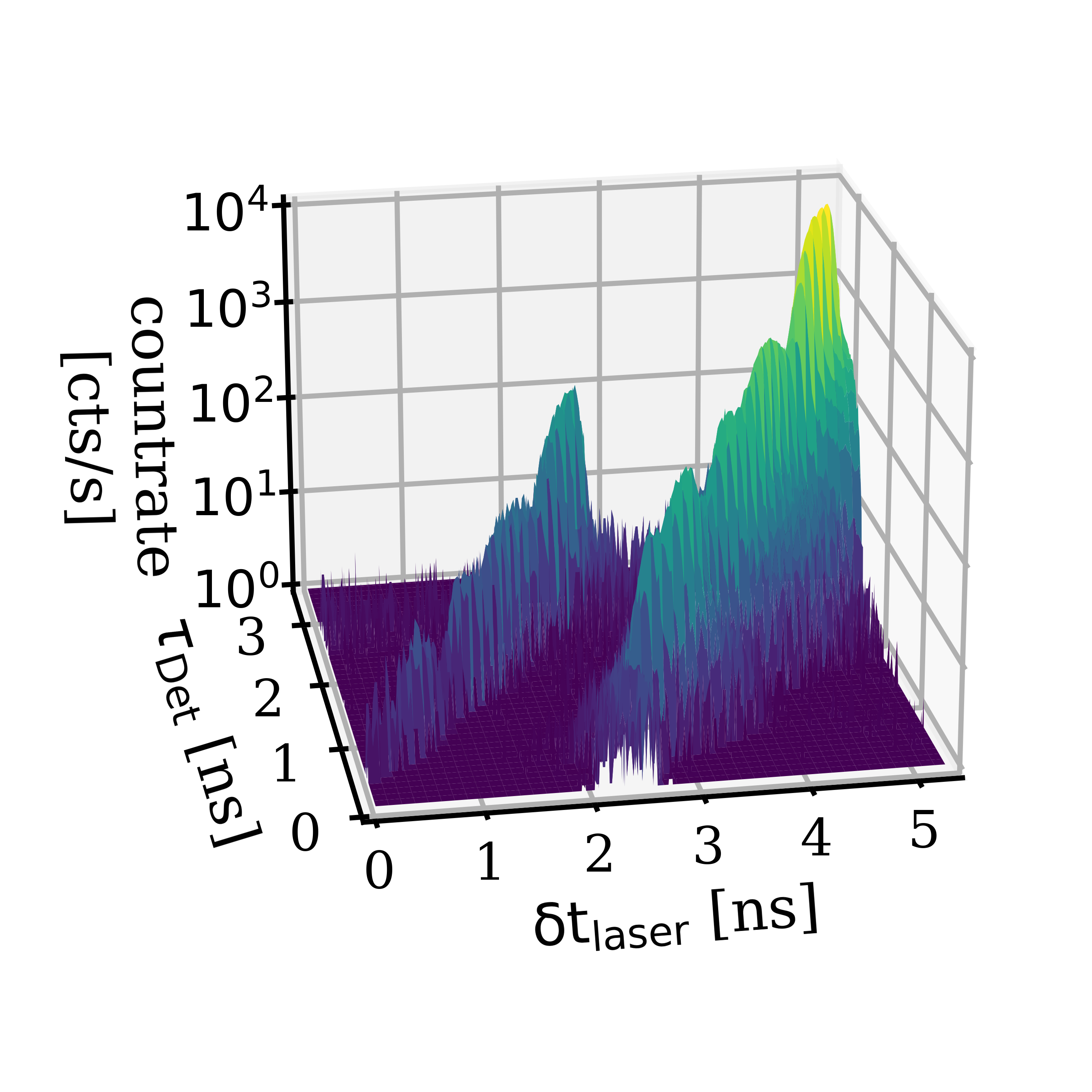}
\caption{Pump-probe emulation where the second pulse is delayed by $2.8\,\mathrm{ns}$ and attenuated by $15\,\mathrm{dB}$ compared to the first pulse.
The horizontal axis shows the time delay $\mathrm{\Phi}$, the vertical axis the time difference between the gate-trigger and SNSPD-signal, and the Z-axis shows the detection count rate of the SNSPD.
The bright pulse is extinguished by $31\pm2\,\mathrm{dB}$ when the gate is turned on between the two pulses.
}
\label{Fig:App2Phot}
\end{figure}

Temporal filtering is based on the principle that the detector is turned on or turned off between the arrival of the pump and signal light pulses.
A proof of principle pump-probe emulation experiment is performed with an attenuated pulsed laser which is split in two pulses with the second pulse delayed by $2.8\,\mathrm{ns}$ and attenuated by $15\,\mathrm{dB}$ compared to the first pulse.
Since the $50\%$ recovery time of the SNSPD is extracted as $3.0\pm0.1\,\mathrm{ns}$, indicating that a DC operation would induce at least $50\%$ loss in the second pulse.
Added to the loss in detection efficiency for the signal pulses, the pump pulse increase the detection rate with the TDC.
The SNSPD is operated in gated mode at the shortest gate window ($5.0\,\mathrm{ns}$ pulse width and $2.4\,\mathrm{ns}$ rise time) at a maximum bias current of $\mathrm{I_b}=19.18\pm0.07\,\mathrm{ \mu A}$.
The variable time delay of the laser $\mathrm{\delta t_{laser}}$ is used to delay the two different optical pulses through the gate window, such that either one of them is at the maximum detection efficiency of the detector, or that both are at a rising or falling flank of the detection efficiency.
Due to the configuration a short time delay ensures that the bright pulse occur before the detector is fully turned on while the dim pulse is at the maximum detection efficiency.
Increasing the time delay $\mathrm{\delta t_{laser}}$ moves the dim pulse through the current unloading edge, while the bright pulse moves through current loading edge toward the maximum detection efficiency.
Histograms of the detection events relative to the gate trigger are measured for different laser delays $\mathrm{\delta t_{laser}}$ and shown in Fig.~\ref{Fig:App2Phot} as a $3$D-map of the pump-probe emulation.
The horizontal axis is the time delay $\mathrm{\delta t_{laser}}$, the vertical axis \red{is} the time difference between the gate-trigger and SNSPD-signal $\tau_\mathrm{Det}$ while the Z-axis shows the count rate on a logarithmic scale.
Scanning the two photon pulses ($500\,\mathrm{fs}$) through the gate window (increasing $\mathrm{\delta t_{laser}}$) shows that the gate window can selectively filter out one of the two pulses while the other optical pulse is detected by the SNSPD.
The bright pulse count rate contribution is extinguished by $31\pm2\,\mathrm{dB}$ when the dim second pulse is at maximum detection efficiency.
Temporal filtering of an optical pulse can be used in experiments requiring a strong laser pulse to generate a signal and where the laser pulse require extinction, such as lifetime measurements or parametric down conversion sources.

\section{Conclusion}
In this work we introduced a method for active electrical gating of an SNSPD.
We demonstrated gating windows from $50\%$ duty cycle down to $5.0\pm0.2\,\mathrm{ns}$ with an off to on rise time of $2.4\pm0.1\,\mathrm{ns}$.
The gated operation mode is simulated and modelled to investigate limitations on the current loading time with which we extracted that our used SNSPD has an upper limit to the inductance of $260\,\mathrm{nH}$.
The characterisation of this SNSPD in three different operation modes, the DC bias, $500\,\mathrm{ns}$ gate window, and $5\,\mathrm{ns}$ gate window showed no difference in photon detection rates at common bias currents.
Reduced bias currents increased the background counts due to the current loading voltage and parasitic leak through coupling.
The timing jitter of an SNSPD operated in gated mode is in principle the same to the SNSPD with a DC bias, however, at a $5\,\mathrm{ns}$ gating window the photon detection events are close to the rising and falling edge of the gating window, which amplifies any timing jitter between the photon and gate window.
This resulted in an increase of the timing jitter by $29.1\pm0.6\,\mathrm{ps}$ to $\mathrm{\tau_{SG}}=144.31\pm0.41\,\mathrm{ps}$.

Gating of an SNSPD opens up different possibilities such as temporal filtering of optical signals and high dynamic range photon counting.
We showed proof of principle experiments for a pump-probe experiment where we could temporally filter an optical pulse separated by $2.8\,\mathrm{ns}$ by $31\pm2\,\mathrm{dB}$.
The high dynamic range experiment with our detector showed an increase of the maximum count rate by a factor of $11.2$, showing that the dynamic range of an SNSPD can be enhanced by gating.

The results shown in the deterministic gating are encouraging, however, improvements are required to unlock the full potential of this work.
The SNSPD used in this work enabled a current loading time of $2.4\,\mathrm{ns}$ and an increase in timing jitter of approximately $30\,\mathrm{ps}$, the state of the art SNSPDs have lower inductance enabling even lower rise times \cite{Chiles2020,Cherednichenko2021} and lower timing jitter \cite{korzh2020}.
Increasing the maximum detection rate is currently done by multiplexing multiple detectors, allowing count rates in the order of $\mathrm{Gcts/s}$~\cite{Huang2018, Polakovic2020, Perrenoud2021}.
To extract these count rates with a single SNSPD via the high dynamic range counting requires further improvements of the SNSPD and gating circuitry.

Gating of an SNSPD would also benefit from improved electronic control such as better isolation between different elements, shielding from external noise sources, and reduction of parasitic inductance and capacitance.
This would improve the current loading time for SNSPDs with lower inductance.
The parasitic electrical effects could be reduced by fabricating the gating circuitry into a single CMOS IC which would also allow bonding directly to the SNSPD \cite{Widhalm2018}.
\red{Direct integration allows further optimisation of the heatload, allowing lower heatloads per SNSPD and thus enhancing the scalability of the gating circuit opening up to multi-pixel arrays~\cite{Allman2015,Wollman2019}.
}
Finally, the optical input is coupled via flood illumination which strongly limits coupling efficiency.
\red{Low inductance SNSPDs also intrinsically suffer from lower coupling efficiencies due to the decreased covered surface area, but this coupling efficiency can be enhanced by a photonic crystal cavity~\cite{Vetter2016} or specifically designed lens systems~\cite{Zhang2015a}.
This} efficient fibre to SNSPD coupling is essential for practical applications with gated SNSPDs such as pump filtering and high dynamic range photon counting.

\begin{appendix}
\section{Acknowledgements}
The authors would like to acknowledge Timon Schapeler and Frederik Thiele for the fruitful discussions.
\section{Funding}
This research is funded by the Bundesministerium für Bildung und Forschung (13N14911 and 3N15856), and by the Deutsche Forschungsgemeinschaft (DFG, German Research Foundation) - Projektnummer No.~231447078–TRR~142.
\section{Disclosures}
A. F. is employed by Single Quantum and might benefit financially.
Other authors declare no conflict of interest.
\section{Data Availability}
Data underlying the results presented in this paper are not publicly available at this time but may be obtained from the authors upon reasonable request.
\section{Supplementary Materials}
See Supplement 1 for supporting content
\end{appendix}



\bibliographystyle{ieeetr}
\bibliography{Bibliography_Gating.bib}

\end{document}


\maketitle


\section{Modelling the current loading time}
\begin{figure}[b!]
\centering\includegraphics{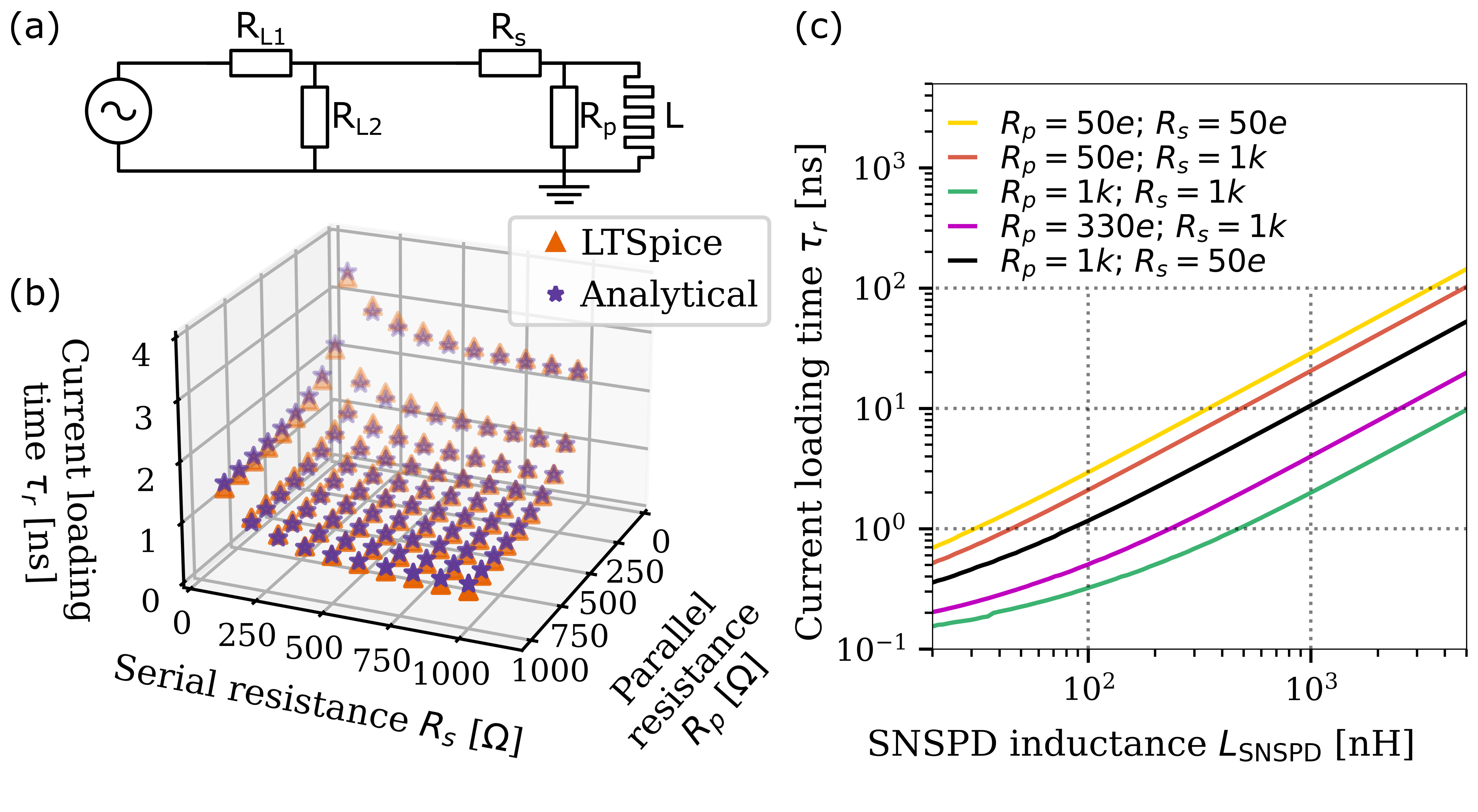}
\caption{Modelling of the current loading times on the SNSPD with varying inductance, parallel resistor, and serial resistor.
(a) The attenuator circuit is approximated by a voltage divider created with $\mathrm{R_{L1}}$ and $\mathrm{R_{L2}}$.
The full attenuator circuit is modelled in LTspice and the extracted current loading time is plotted with a fixed inductance ($220\,\mathrm{nH}$) and varying parallel and serial resistor is shown in (b) with the orange triangles.
The current loading times as calculated with Eq. \ref{EQ:SupLoadTime} is shown as purple stars, where $\mathrm{R_{L1}}$ and $\mathrm{R_{L2}}$ are fitting parameters.
(c) The effect of the inductance on the current loading time for five different resistor configurations simulated with the LTspice model.
The cut-off in the loading time comes from the $300\,\mathrm{ ps}$ rise time limit of the inverter generating the voltage pulse.
}
\label{Fig:SupElecModel}
\end{figure}
The timing characteristics of the current loading and unloading through the SNSPD depends on the full electrical circuit controlling the bias current.
To investigate the behaviour of the bias current (un)loading response we simulate the full attenuation circuitry in LTspice~\cite{LTSpiceProg2020}, which is then compared to an analytical model with a simplified version of the attenuation circuit.
The extraction from LTspice is performed by simulating every configuration of $\mathrm{R_{s}}$, $\mathrm{R_{p}}$, and $\mathrm{L}$.
The attenuation circuit fixed at the design parameters from the physical attenuator circuit used for the gating measurements.
A $300\,\mathrm{ps}$ rise time is used for the inverter, since this is the measured fastest rise time at room temperature of this inverter type.
The time trace for every configuration is then analysed to obtain the $10\%-90\%$ rise time on the current loading through the inductor, which is transformed to a exponential time constant for comparison to the analytical model.

The analytical result of the simplified attenuation circuit allows insight in the effect of relevant parameters on the current (un)loading time through the analytical expression.
The model of the simplified electrical circuit is shown in Fig.~\ref{Fig:SupElecModel}~(a) where the attenuation circuit is replaced by a voltage divider, and the bias voltage and generated pulses are replaced by a function generator.
\red{The SNSPD is replaced as a pure inductor since we assume that the current only loads when the SNSPD is superconductive (and thus $R_{SNSPD}=0\,\Omega$).
Residual parasitic resistance and capacitance can be expected, the order of magnitude is however expected to be negligible and thus not modelled}.
The current loading behaviour in this model is calculated by applying a step response as the bias voltage from which an (inverted) exponential decay is derived as the current loading shape.
The time constant of this exponential decay is extracted as 
\begin{equation}
\tau_{\mathrm{rm}} =
 \frac{
 \mathrm{L}\left(
 \mathrm{R_{s}}\left(2\mathrm{R_{L1}+R_{L2}}\right)+\mathrm{R_{p}\left(R_{L1}+R_{L2}\right)}
 \right)}
 {\mathrm{R_p R_s}\left(2\mathrm{R_{L1}+R_{L2}}\right)}~,
 \label{EQ:SupLoadTime}
\end{equation}
where $\mathrm{R_{L1}}$ and $\mathrm{R_{L2}}$ are the substitute resistors for the attenuator, $\mathrm{R_p}$ and $\mathrm{R_s}$ are respectively the parallel and serial resistors, and $\mathrm{L}$ is the SNSPD inductance.

This shows that the current loading time responds differently to changes in $\mathrm{R_s}$ than to changes in $\mathrm{R_p}$ where the difference in behaviour is determined by the attenuation circuit.
The LTspice simulations do not use a step function as the input voltage but a ramp with a rise time of $300\,\mathrm{ps}$.
To compensate for this difference, the rise time $\tau_\mathrm{{rm}}$ from the model is lower-bounded with a minimum rise time $\tau_{\mathrm{Min}}$ via $\tau_r = \sqrt{\tau_{\mathrm{rm}}^2+\tau_{\mathrm{Min}}^2}$.

The validity of this model is tested by fitting the obtained rise time from the model to the actual simulation of the full circuitry in LTspice, with fitting parameters $\tau_{\mathrm{Min}}$, $\mathrm{R_{L1}}$, and $\mathrm{R_{L2}}$.
Figure~\ref{Fig:SupElecModel}~(b) compares the LTspice simulation (orange triangles) and the fit of the simplified model (purple starts) for a fixed inductance of $220\,\mathrm{nH}$ and varying $\mathrm{R_s}$ and $\mathrm{R_p}$.
The extracted fitting parameters for $\mathrm{R_{L1}}$ and $\mathrm{R_{L2}}$ are in the order of $10^7\,\mathrm{\Omega}$, indicating a clear decoupling between the input voltage and the current through $\mathrm{R_{s}}$ and justifying the simplification of the attenuator.
The extracted minimum rise time $\tau_{\mathrm{Min}}\approx550\,\mathrm{ns}$.
The proximity between the simulation and model shows another validation for the approximations used to derive the current loading time.

From Eq.~\ref{EQ:SupLoadTime} and Fig.~\ref{Fig:SupElecModel}~(b) it can be derived that the current loading time is reduced with an increase in the parallel and serial resistor, the effect of the serial resistor is however smaller than the parallel resistor.

Equation~\ref{EQ:SupLoadTime} indicates a linear behaviour between the current loading time and the inductance of the SNSPD.
This behaviour is also simulated with LTspice and the current loading time depending on the inductance of the SNSPD is displayed in Fig.~\ref{Fig:SupElecModel}~(c) for different combinations of the parallel and serial resistor.
The simulation shows a predominantly linear relationship between the inductance and current loading time, before becoming saturated when the current loading time starts to become limited by the influence of the parasitic elements of the components.
\bibliographystyle{ieeetr}
\bibliography{Bibliography_Gating.bib}